\documentclass[12pt]{article}
\usepackage{amsthm,amsmath,amssymb,amsfonts}
\usepackage{graphicx,psfrag,epsf}
\usepackage{enumerate}
\usepackage[round]{natbib}
\usepackage{bm}
\usepackage[titletoc]{appendix}
\usepackage{color}
\usepackage{xcolor}
\usepackage{multirow}
\usepackage{url}
\usepackage{animate}
\usepackage{graphicx,geometry}
\usepackage{changes}
\usepackage{ulem}
\usepackage{authblk}
\usepackage{adjustbox}
\fontfamily{times}

\usepackage{booktabs}
\usepackage[skip=0.5\baselineskip,font=small]{caption}
\usepackage{float}
\usepackage{etoolbox}
\usepackage{indentfirst}
\usepackage{enumitem}
\usepackage{threeparttable}
\usepackage[unicode,colorlinks=true,linkcolor=black,urlcolor=blue]{hyperref}
\usepackage{subcaption}
\usepackage{makecell}
\usepackage{array}
\usepackage{multirow,caption,adjustbox,float}
\usepackage{setspace}

\bmdefine{\bgamma}{\gamma}
\bmdefine{\bzeta}{\zeta}

\newcommand{\argmin}{\operatornamewithlimits{argmin}}

\date{\today}

\title{\bf DeepKriging on the global Data}
\author{\small{
Hao-Yun Huang$^{1}$,
Wen-Ting Wang$^{1,2}$,
Ping-Hsun Chiang$^{1}$,
Wei-Ying Wu$^{1}$
}
}

\affil{\small{
$^{1}$Department of Applied Mathematics, National Dong Hwa University\\
$^{2}$Department of Applied Mathematics, National Chung Hsing University\\
Emails: hhuscout@gms.ndhu.edu.tw; egpivo@gmail.com; andrew501228@gmail.com; wuweiying1011@gms.ndhu.edu.tw
}}

\begin{document}
	\baselineskip=18pt
	\maketitle

\begin{abstract}
The increasing availability of large-scale global datasets has generated a demand for scalable spatial prediction methods defined on spherical domains. Classical spatial models that rely on Euclidean distance representations are inappropriate for spherical data because planar projections distort geodesic distances and spatial neighborhood structures, while traditional kriging-based prediction methods are often computationally prohibitive for massive datasets. To address these challenges, we propose a Spherical DeepKriging framework for spatial prediction on $\mathbb{S}^2$. The proposed approach constructs a flexible prediction model by integrating thin-plate spline (TPS) basis functions defined intrinsically on the sphere. Simulation studies and real data analyses are presented to demonstrate the superior predictive performance of the proposed method.

\end{abstract}

\noindent keywords: Spherical domains; Thin-plate splines; Kriging; Deep learning

\section{Introduction}
Spatial prediction is a fundamental problem in spatial statistics, concerned with 
inferring an underlying spatial process at unobserved locations based on observations 
collected at a finite set of sites. It plays a central role in geostatistics and has 
broad applications in environmental monitoring, climate science, remote sensing, and 
epidemiology. Classical spatial prediction methods are predominantly kriging-based 
approaches, which provide best linear unbiased prediction (BLUP) under Gaussian 
assumptions \citep{cressie1993statistics, stein1999interpolation, diggle2007model, 
banerjee2014hierarchical, rasmussen2006gaussian}. Despite their optimality properties, 
these methods become computationally infeasible for massive datasets due to their 
$O(n^3)$ computational complexity and numerical instability associated with large 
covariance matrix inversions, where $n$ denotes the sample size.

To mitigate these computational challenges, a substantial literature has developed 
approximate and reduced-rank approaches, including covariance tapering, Vecchia and 
nearest-neighbor approximations, and multi-resolution constructions 
\citep{furrer2006tapering, vecchia1988estimation, datta2016nngp, katzfuss2017multi}. 
Although these approaches improve scalability, they often require careful model design 
and may sacrifice predictive flexibility.

Recently, DeepKriging methods have emerged as a scalable alternative that integrates 
spatial statistical modeling with deep learning. Let observations from a spatial 
process be collected at locations $\{\mathbf{s}_i\}_{i=1}^n$ in a domain $D$, and assume
\begin{equation}
z(\mathbf{s}_i) = y(\mathbf{s}_i) + \varepsilon(\mathbf{s}_i), 
\qquad i = 1, \ldots, n,
\label{eq:measurement_dk}
\end{equation}
where $y(\mathbf{s})$ denotes the spatial process of interest and 
$\{\varepsilon(\mathbf{s}_i)\}_{i=1}^n$ are independent measurement errors. The goal of 
spatial prediction is to construct an accurate predictor of $y(s_0)$ at an unobserved 
location $s_0 \in D$.

Within the DeepKriging framework, this task is formulated as a supervised learning 
problem, in which the predictor is approximated by a neural network (NN) through 
minimization of an expected loss function,
\begin{equation}
\hat{f}_{\mathrm{NN}}(s_0)
:= \argmin_{\theta \in \Theta} 
\mathbb{E}\!\left\{ \mathcal{L}\!\left(y(s_0), \hat f_{\theta}(s_0)\right) \right\},
\end{equation}
where $\hat f_{\theta}$ denotes the possible model introduced through NN with parameter space $\Theta$ and 
$\mathcal{L}(\cdot,\cdot)$ is a loss function chosen according to the learning task. In 
practice, the parameters $\theta$ are estimated by empirical risk minimization,
\begin{equation}
\hat{\theta}
:= \argmin_{\theta \in \Theta} 
\frac{1}{n} \sum_{i=1}^{n} 
\mathcal{L}\!\left(z_i, \hat f_{\theta}(\mathbf{s}_i)\right),
\end{equation}
leading to the predictor $\hat{y}_{\hat{\theta}}(s_0):=\hat f_{\hat{\theta}}(s_0)$.

A key component of DeepKriging is the construction of informative input features. 
Rather than using raw spatial coordinates, DeepKriging employs spatial basis functions 
to encode spatial dependence within the NN input layer. Specifically, the input feature vector at location $\mathbf{s}$ is defined as
{
\begin{equation}
\label{eq:input_feature}
\mathbf{u}(\mathbf{s})
=\big(\boldsymbol{\phi}(\mathbf{s})^{\top},\ \mathbf{x}(\mathbf{s})^{\top}\big)^{\top},
\end{equation}
}
where {$\mathbf{x}(\mathbf{s})$} denotes observed covariates and {$ {\boldsymbol{\phi}}(\mathbf{s})$} denotes a collection of spatial basis functions.

Recent studies have explored different choices of basis functions within this 
framework. \citet{chen2024deepkriging} employed Wendland basis functions, which are closely related to fixed-rank kriging covariance 
structures. However, Wendland bases are compactly 
supported and radial, often requiring a large number of basis functions to adequately 
cover the spatial domain. When observations are highly irregularly spaced, some basis 
supports may contain few or no data points, potentially leading to numerical instability 
during training. To address these issues, \citet{lin2023some} introduced multi-resolution 
thin-plate spline (MRTS) basis functions \citep{tzeng2018resolution} within a low-rank 
thin-plate spline framework and further incorporated Huber’s loss \citep{Huber1992} to 
improve robustness.

Nevertheless, existing DeepKriging methods based on Wendland or MRTS bases are primarily 
developed for Euclidean domains. In many scientific applications, including global 
climate studies and astronomy, data are collected on spherical surfaces. When the 
observation domain is spherical, Euclidean basis constructions do not intrinsically 
respect the underlying geometry, which may result in geometric distortion and degraded 
predictive performance. These considerations motivate the development of DeepKriging 
methodologies that are intrinsically defined on spherical domains.

In this paper, inspired by \citet{chen2024deepkriging}  and \citet{lin2023some},  we consider a DeepKriging framework based on spherical basis functions 
defined directly on the sphere $\mathbb{S}^2$. Through simulation studies and real data 
analyses, we investigate how the choice of spatial basis functions influences 
prediction performance under appropriate learning settings. The remainder of the paper 
is organized as follows. Section \ref{sec:Method} introduces the proposed spherical basis 
construction and the associated DeepKriging model, while Section \ref{sec:num}
 presents numerical 
studies to evaluate its predictive performance. The implementation is publicly available at https://github.com/STLABTW/spherical-deepkriging.


\section{Methodology}\label{sec:Method}
Assume that the spatial data $z(\mathbf{s})$ are defined at locations $\mathbf{s}$ on the unit sphere $\mathbb{S}^2$, with spherical coordinates $(\lambda,\theta)^\top \in [0,2\pi)\times[0,\pi)$. Suppose we collect observations at $n$ such locations, denoted by $\{\mathbf{s}_i\}_{i=1}^n \subset \mathbb{S}^2$, according to the measurement model in \eqref{eq:measurement_dk}. For notational convenience, we write $z_i=z(\mathbf{s}_i)$ and $y_i=y(\mathbf{s}_i)$ for $i=1,\ldots,n$.

To accommodate global data observed on $\mathbb{S}^2$, we use spherical basis functions as input features to a neural network predictor. In particular, we first review the spherical multi-resolution thin-plate spline basis system (Spherical MRTS) proposed by \citet{Huang2025} and then describe the corresponding DeepKriging architecture and training procedure.

\subsection{Spherical multi-resolution 
thin-plate spline basis functions (Spherical MRTS)\label{subsec:Spherical MRTS}} 

\paragraph{Spherical TPS roughness penalty}
Let $\psi(\mathbf{s})$ be a real-valued function on $\mathbb{S}^2$. The spherical TPS formulation is motivated by the penalized criterion
\begin{equation}
  \min_{\psi \in \mathcal{H}_2(\mathbb{S}^2)}
  \ \sum_{i=1}^{n}\Big\{ z(\mathbf{s}_i) - \psi(\mathbf{s}_i) \Big\}^{2}
  + \rho\, J_{2}(\psi),
  \label{eq:tps_sphere_obj}
\end{equation}
 where $\rho>0$ is a smoothing parameter and $J_2(\psi)$ is the roughness penalty defined via the Laplace--Beltrami operator on $\mathbb{S}^2$:
\begin{equation}
J_{2}(\psi)
= \int_{0}^{2\pi}\!\!\int_{0}^{\pi}
  \big( \Delta \psi(\lambda, \theta) \big)^{2}
  \sin \theta \, d\theta \, d\lambda,
\quad
\Delta \psi
= \frac{1}{\sin \theta}\frac{\partial}{\partial \theta}
  \left( \sin \theta \frac{\partial \psi}{\partial \theta} \right)
  + \frac{1}{(\sin \theta)^2}\frac{\partial^{2}\psi}{\partial \lambda^{2}}.
\label{eq:roughness_lb}
\end{equation}
This penalty enforces smoothness with respect to the intrinsic geometry of $\mathbb{S}^2$ and leads to a kernel-based representation of the solution.

Following \citet{beatson2018thinplate}, the spherical TPS kernel is
\begin{equation}
\Phi(\mathbf{s},\mathbf{s}^{\ast})
= \mathrm{Li}_2\!\left(\tfrac{1}{2}+\tfrac{\cos(\gamma(\mathbf{s},\mathbf{s}^{\ast}))}{2}\right)
  + 1 - \tfrac{\pi}{6},
\label{eq:spherical_tps_kernel}
\end{equation}
where $\mathrm{Li}_2(y)=-\!\int_0^y\frac{\log(1-x)}{x}\,dx$ is the dilogarithm function. The kernel $\Phi(\cdot,\cdot)$ depends on the great-arc angle $\gamma(\mathbf{s},\mathbf{s}^{\ast})$ between $\mathbf{s}$ and $\mathbf{s}^{\ast}$.

\paragraph{Knots and eigen-ordered multi-resolution bases}
To obtain a computationally efficient basis representation, we construct the spline bases using a set of $m$ control points (knots number)
$\{\boldsymbol{\kappa}_1,\ldots,\boldsymbol{\kappa}_m\}\subset\mathbb{S}^2$.
Define the kernel matrix $\bm{K}\in\mathbb{R}^{m\times m}$ by
\begin{equation}
\bm{K}_{jj'}=\Phi(\boldsymbol{\kappa}_j,\boldsymbol{\kappa}_{j'}),
\qquad j,j'=1,\ldots,m,
\end{equation}
and the centering matrix
\begin{equation}
\bm{Q}=\bm{I}_m-\bm{1}_m\bm{1}_m^{\top}/m,
\end{equation}
where $\bm{1}_m$ is the $m$-vector of ones. Consider the eigendecomposition
\begin{equation}
\bm{Q}\bm{K}\bm{Q}=\bm{V}\bm{\Lambda}\bm{V}^{\top},
\qquad
\Lambda_1\ge \Lambda_2 \ge \cdots \ge \Lambda_m = 0,
\label{eq:eig_spherical}
\end{equation}
where $\bm{V}=(\bm{v}_1,\ldots,\bm{v}_m)$ and $\bm{\Lambda}=\mathrm{diag}(\Lambda_1,\ldots,\Lambda_m)$.
For any $\mathbf{s}\in\mathbb{S}^2$, define the kernel vector with respect to the knots:
\begin{equation}
\bm{k}(\mathbf{s})
=\big(\Phi(\mathbf{s},\boldsymbol{\kappa}_1),\ldots,\Phi(\mathbf{s},\boldsymbol{\kappa}_m)\big)^{\top}\in\mathbb{R}^m.
\end{equation}
We then construct an orthogonal family of spherical spline basis functions $\{\phi_k(\mathbf{s})\}_{k=1}^{m}$ ordered by smoothness:
\begin{equation}
\phi_k(\mathbf{s}) =
\begin{cases}
m^{-1/2}, & k = 1, \\[2mm]
\Lambda_{k-1}^{-1}
\big( \bm{k}(\mathbf{s}) - \bm{K}\bm{1}_m / m \big)^{\!\top} \bm{v}_{k-1},
& k = 2,\ldots,m.
\end{cases}
\label{eq:mr_spherical_basis}
\end{equation}
This eigen-ordered construction yields a natural multi-resolution representation: components associated with larger eigenvalues tend to be smoother (capturing large-scale variation), whereas smaller eigenvalues correspond to less smooth components (capturing finer-scale variation). Moreover, the roughness penalty satisfies $J_2(\phi_k)=\Lambda_{k-1}^{-1}$ for $k=2,\ldots,m$, linking the basis ordering directly to smoothness.

\subsection{Neural Network Architecture and Training \label{subsec:NN}}
\begin{figure}[t]
  \centering
  \includegraphics[width=0.45\textwidth]{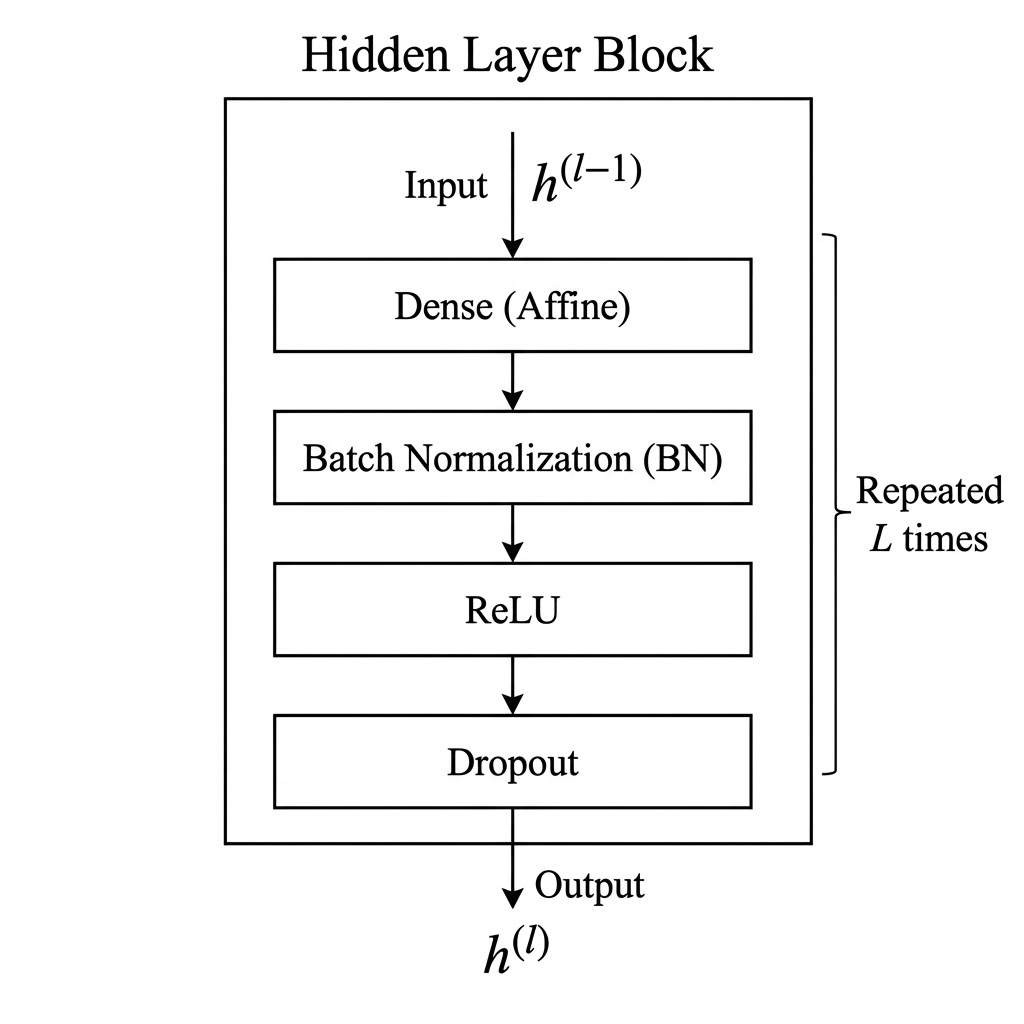}
  \caption{Structure of a hidden-layer block in the DeepKriging neural network.
  Each block consists of an affine transformation followed by batch normalization,
  ReLU activation, and dropout, and is repeated $L$ times.}
  \label{fig:dk_block}
\end{figure}

{
Let $\mathcal{I}_{\mathrm{train}}$ and $\mathcal{I}_{\mathrm{val}}$ denote the index sets of the training and validation samples, respectively, with $\mathcal{I}_{\mathrm{train}}, \mathcal{I}_{\mathrm{val}} \subseteq \{1,\ldots,n\}$ and $\mathcal{I}_{\mathrm{train}} \cap \mathcal{I}_{\mathrm{val}} = \emptyset$. For each location $\mathbf{s}$, we construct the input feature vector $\mathbf{u}(\mathbf{s})$ as defined in \eqref{eq:input_feature}, where $\boldsymbol{\phi}(\mathbf{s})=(\phi_1(\mathbf{s}),\ldots,\phi_K(\mathbf{s}))^\top$ is the $K$-dimensional spherical MRTS basis feature vector given in \eqref{eq:mr_spherical_basis}, and $\mathbf{x}(\mathbf{s})$ denotes additional covariates (if available). We use $\mathbf{u}(\mathbf{s})$ as the input-layer representation, i.e., $\mathbf{h}^{(0)}(\mathbf{s})=\mathbf{u}(\mathbf{s})$.

The DeepKriging predictor is implemented as a fully connected multilayer perceptron (MLP) with $L$ hidden blocks (Figure~\ref{fig:dk_block}). Each block applies an affine transformation followed by batch normalization (BN) and an activation function $\sigma(\cdot)$; dropout is applied after the activation during training as a regularization step \citep{cai2019effective}. In our implementation, we set $\sigma(\cdot)=\mathrm{ReLU}(\cdot)$. For $l=1,\ldots,L$,
\begin{equation}
\label{eq:learning}
\mathbf{h}^{(l)}(\mathbf{s})
=
\sigma\!\left(
\mathrm{BN}^{(l)}\!\left(\mathbf{W}^{(l)}\mathbf{h}^{(l-1)}(\mathbf{s})+\mathbf{b}^{(l)}\right)
\right),
\end{equation}
where $\mathbf{W}^{(l)} \in \mathbb{R}^{d_l \times d_{l-1}}$ and $\mathbf{b}^{(l)} \in \mathbb{R}^{d_l}$ are trainable parameters, $d_l$ denotes the width of the $l$-th hidden layer, and $\mathrm{BN}^{(l)}(\cdot)$ denotes batch normalization with layer-specific trainable scale and shift parameters. The output layer is linear:
\begin{equation}
\label{eq:output_layer}
\hat y_{\theta}(\mathbf{s})=\mathbf{W}^{(L+1)}\mathbf{h}^{(L)}(\mathbf{s})+\mathbf{b}^{(L+1)}.
\end{equation}
Note that $\theta$ collects all trainable parameters, including $\{\mathbf{W}^{(l)},\mathbf{b}^{(l)}\}_{l=1}^{L+1}$ and the batch-normalization parameters. We write $\theta(K)$ to emphasize that the parameterization depends on $K$ through the input dimension.
}

{
Given the training data $\{(\mathbf{u}_i,z_i)\}_{i\in\mathcal{I}_{\mathrm{train}}}$ with $\mathbf{u}_i=\mathbf{u}(\mathbf{s}_i)$, we minimize the empirical training risk
\begin{equation}
\label{eq:train_obj}
\mathcal{R}_{\mathrm{train}}\!\big(K;\theta\big)
=
\frac{1}{|\mathcal{I}_{\mathrm{train}}|}
\sum_{i\in\mathcal{I}_{\mathrm{train}}}
\mathcal{L}\!\left(z_i,\hat y_{\theta}(\mathbf{s}_i)\right),
\end{equation}
and use an iterative optimizer to generate a sequence of parameter iterates $\{\theta^{(t)}(K)\}_{t=1}^{T}$.
}


{
We train the network using the Adam optimizer (Adaptive Moment Estimation; \citealp{kingma2015adam}). Because the objective is non-convex, global optimality is not guaranteed. To mitigate overfitting, we apply early stopping based on the validation empirical risk. For a fixed $K$, define
\begin{equation*}
\mathcal{R}_{\mathrm{val}}\!\big(K;\theta^{(t)}(K)\big)
=
\frac{1}{|\mathcal{I}_{\mathrm{val}}|}
\sum_{j \in \mathcal{I}_{\mathrm{val}}}
\mathcal{L}\!\left(
z_j,\ \hat{y}_{\theta^{(t)}(K)}(\mathbf{s}_j)
\right).
\end{equation*}
For each candidate $K$ in a set $\mathcal{K}$, we first identify the iterate attaining the smallest validation risk during training:}
\begin{equation}
\label{eq:val_loss}
\widehat{\theta}(K)\in\arg\min_{\theta \in \{\theta^{(t)}(K)\}_{t=1}^{T}}
\mathcal{R}_{\mathrm{val}}\!\big(K;\theta\big).
\end{equation}
The optimal number of basis functions (or resolutions), denoted by $\widehat{K}$, is selected by minimizing the validation risk over all candidate values:
\begin{equation}
    \label{eq:val_loss2}
    \widehat{K} = \arg\min_{K \in \mathcal{K}} \mathcal{R}_{\mathrm{val}}\!\big(K;\widehat{\theta}(K)\big).
\end{equation}

\section{Simulation Studies \& Real Data Analysis}

To demonstrate the superior performance of the proposed DeepKriging framework based on spherical MRTS basis functions for global data analysis, we compare it with several widely used benchmark approaches, including ordinary least squares (OLS), the universal kriging, and existing DeepKriging methodologies constructed using Euclidean Wendland bases \citep{chen2024deepkriging} and Euclidean MRTS bases \citep{lin2023some}.

In both OLS and universal kriging, we use the augmented feature vector
\begin{equation}
\bm{x}_i=\bigl(\boldsymbol{\phi}(\bm{s}_i)\bigr) \in \mathbb{R}^{K}
\label{eq:design vector}
\end{equation}
as the design vector, where $\boldsymbol{\phi}(\bm{s}_i)$ denotes the spatial basis features.

In the simulation studies, OLS models constructed with Euclidean-Wendland and spherical MRTS bases are denoted by $\mathrm{OLS}_W$ and $\mathrm{OLS}_S$, respectively. Universal kriging is implemented using the spherical MRTS basis and is denoted by $\mathrm{UK}$. For the DeepKriging framework, we consider models based on Wendland, and Euclidean and  spherical MRTS bases, denoted by DK\_W, DK\_MRTS and DK\_S, respectively. 
To enhance the robustness of the proposed method against outliers, we further train the spherical DeepKriging model using the Huber loss \citep{Huber1992} with the threshold parameter set to $\delta=1.345$, denoting the corresponding results as DK\_S\_H.

For each scenario, we perform 50 replications. In each replication, we simulate $n=2500$ samples and randomly split the data into 80\% training set ($\mathcal{I}_{\mathrm{tr}}$), 10\% validation set ($\mathcal{I}_{\mathrm{val}}$), and 10\% testing set ($\mathcal{I}_{\mathrm{test}}$).

The size of $\boldsymbol{\phi}(\cdot)$ depends on the chosen family of basis functions. For the Wendland basis, we adopt commonly used multi-resolution settings with $10^2$, $19^2$, and $37^2$ basis functions across three scales; this configuration is used for both OLS and DeepKriging, where the latter employs a four-layer neural network architecture (i.e., three hidden layers).

For MRTS-type bases, the size of basis functions is selected by minimizing the validation loss defined in \eqref{eq:val_loss2}. In particular, we use the mean squared error for models trained under the squared-error criterion and the Huber loss for models trained under a robust objective.

To quantify predictive performance,
 let $z_i=z(\bm{s}_i)$ and $\widehat{y}_i=\widehat{y}_{\hat{\theta}}(\bm{s}_i)$ respectively denote the true value and the corresponding prediction for $i\in  \mathcal{I}_{\mathrm{test}}.$ 
 We consider 
the Root Mean Squared Error (RMSE) by
\begin{equation}
\mathrm{RMSE}
=
\sqrt{
\frac{1}{|  \mathcal{I}_{\mathrm{test}}|}
\sum_{i\in  \mathcal{I}_{\mathrm{test}}}
\left(z_i-\widehat{y}_i\right)^2},
\end{equation}
the Mean Absolute Error (MAE) by
\begin{equation}
\mathrm{MAE}
=
\frac{1}{|  \mathcal{I}_{\mathrm{test}}|}
\sum_{i\in  \mathcal{I}_{\mathrm{test}}}
\left|z_i-\widehat{y}_i\right|.
\end{equation}

The observation locations, $\mathbf{s} = (\lambda, \theta)^\top \in \mathbb{S}^2$, are uniformly generated on the unit sphere by sampling:
\[
\phi \sim \mathrm{Unif}(0,2\pi),\quad
u \sim \mathrm{Unif}(-1,1),
\]
and setting the longitude to $\lambda = \phi - \pi$ and the latitude to $\theta = \arcsin(u)$. To better reflect the imperfections commonly encountered in real-world data, we consider simulation scenarios beyond the ideal noise-free setting. Specifically, we incorporate two types of contamination in the observation model:
(a) Gaussian measurement noise, $N(0, 0.5^2)$; and 
(b) artificial outliers, generated by randomly selecting $2\%$ of the training observations $z(\mathbf{s})$ and multiplying them by $5$.

\subsection{Simulation}

\paragraph{(i) Stationary processes on $\mathbb{S}^2$}\mbox{}\par

We consider the conventional stationary Gaussian process defined by 
\label{sec:num}
\begin{equation}
    y(\bm{s}) \;=\; 1+g(\bm{s}),
\end{equation}
where $"1"$ represents the deterministic  mean structure, and $g(\cdot)$ is a zero-mean Gaussian process with the exponential correlation with the range parameter is 0.5 and the variance parameter is 1. 

\begin{figure}[H]
    \centering
    \includegraphics[scale=0.36]{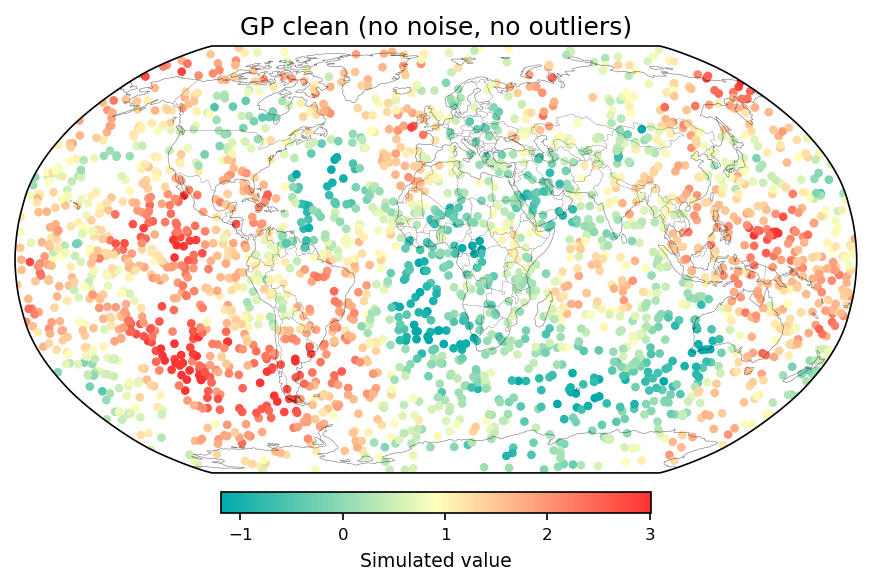}
    \includegraphics[scale=0.36]{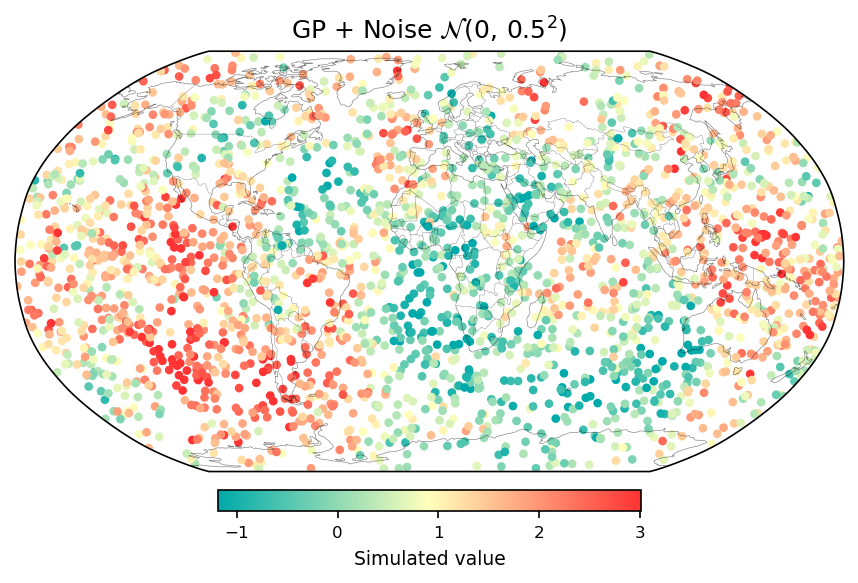}
    \includegraphics[scale=0.36]{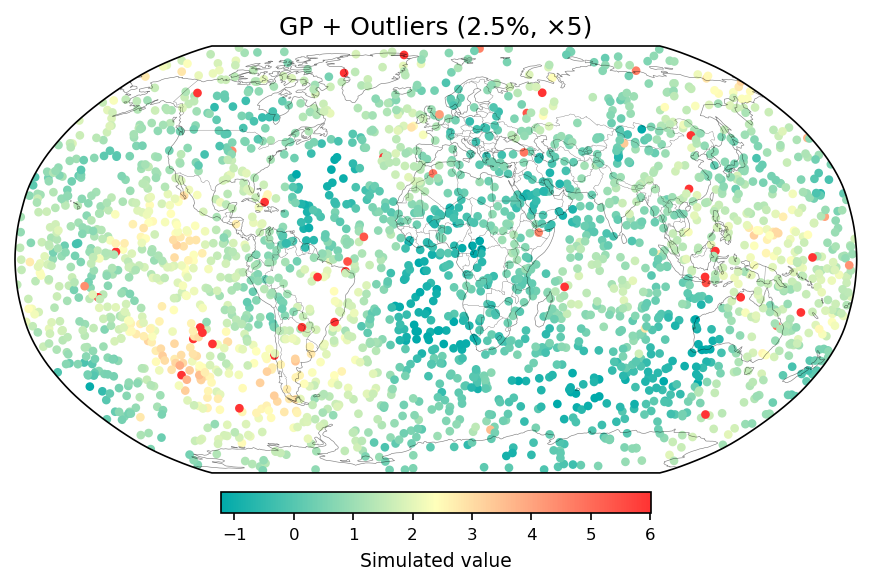}
    \caption{{Realizations of Stationary Gaussian process (i)}}
    \label{fig:sim1}
\end{figure}

The plot of one realization is shown in Figure \ref{fig:sim1}. 
The average of the mean square prediction errors are shown in Tables \ref{tab:m=1clean}-\ref{tab:m=1c=5}, with the corresponding standard deviations given after the plus–minus sign.

\begin{table}[H]
  \centering
  {
  \begin{tabular}{@{} lccc @{}}
    \toprule
    Model & RMSE & MAE \\
    \midrule
    OLS\_W    & \(3.49 \pm 8.39\) & \(0.85 \pm 0.56\) \\
    OLS\_S    & \(0.39 \pm 0.02\) & \(0.31 \pm 0.02\) \\
    DK\_W     & \(0.71 \pm 0.09\) & \(0.54 \pm 0.07\) \\
    DK\_MRTS  & \(0.39 \pm 0.03\) & \(0.31 \pm 0.02\) \\
    DK\_S     & \(0.32 \pm 0.02\) & \(0.25 \pm 0.01\) \\
    DK\_S\_H  & \(0.32 \pm 0.02\) & \(0.25 \pm 0.01\) \\
    UK        & \(0.30 \pm 0.01\) & \(0.24 \pm 0.01\) \\
    \bottomrule
  \end{tabular}
  }
  \caption{{Prediction performance for  $m(x)=1$ without $\varepsilon(\bm{s})$.}}
  \label{tab:m=1clean}
\end{table}

\begin{table}[H]
  \centering
  \small
  {
  \begin{tabular}{@{} lcc @{}}
    \toprule
    Model & RMSE & MAE \\
    \midrule
    OLS\_W    & $4.79 \pm 9.76$ & $1.08 \pm 0.67$ \\
    OLS\_S    & $0.66 \pm 0.03$ & $0.53 \pm 0.03$ \\
    DK\_W     & $0.89 \pm 0.08$ & $0.70 \pm 0.06$ \\
    DK\_MRTS  & $0.67 \pm 0.04$ & $0.54 \pm 0.03$ \\
    DK\_S     & $0.67 \pm 0.03$ & $0.54 \pm 0.03$ \\
    DK\_S\_H  & $0.68 \pm 0.03$ & $0.54 \pm 0.03$ \\
    UK        & $0.63 \pm 0.03$ & $0.50 \pm 0.03$ \\
    \bottomrule
  \end{tabular}
  }
  \caption{{Prediction performance for $m(x)=1$  with $\varepsilon(\bm{s})\sim N(0,0.5^2)$}}
  \label{tab:noise_normalnoise}
\end{table}

\begin{table}[H]
  \centering
  {
  \begin{tabular}{@{} lcc @{}}
    \toprule
    Model & RMSE & MAE \\
    \midrule
    OLS\_W    & \(4.87 \pm 9.43\) & \(1.08 \pm 0.66\) \\
    OLS\_S    & \(0.94 \pm 0.25\) & \(0.49 \pm 0.07\) \\
    DK\_W     & \(1.14 \pm 0.25\) & \(0.69 \pm 0.09\) \\
    DK\_MRTS  & \(0.98 \pm 0.25\) & \(0.52 \pm 0.07\) \\
    DK\_S     & \(0.98 \pm 0.25\) & \(0.48 \pm 0.07\) \\
    DK\_S\_H  & \(0.91 \pm 0.27\) & \(0.42 \pm 0.06\) \\
    UK        & \(0.91 \pm 0.26\) & \(0.44 \pm 0.06\) \\
    \bottomrule
  \end{tabular}
  }
  \caption{{Prediction performance for $m(x)=1$ with the artificial outliers.}}
  \label{tab:m=1c=5}
\end{table}

\noindent {\textbf{(ii) Local Extremes}\vspace{0.2cm}\par

{To illustrate the approximation capability of spatial basis functions for deterministic structures, we construct a non-smooth deterministic function $m(\mathbf{s})$. We first define a macroscopic base trend featuring a global baseline, sharp exponential peaks at $\pm 45^\circ$ latitude, an equatorial drop, and a non-differentiable block drop representing a mountain effect:
$$
\begin{aligned}
f_{macro}(\lambda, \theta) =\ & 5 + 18 \exp\left(-\frac{(\theta - \pi/4)^2}{0.05}\right) \\
& + 22 \exp\left(-\frac{(\theta + \pi/4)^2}{0.04}\right) - 4 \exp\left(-\frac{\theta^2}{0.01}\right) \\
& - 12 \cdot \mathbb{I}_{\{\lambda \in (0, 1), \theta \in (0.1, 1)\}}
\end{aligned}
$$}

{To inject severe, highly localized topological anomalies, we generate a set of $60$ random spatial extrema, denoted by $\{\mathbf{a}_i\}_{i=1}^{60}$, uniformly sampled on the sphere $\mathbb{S}^2$. At each anomaly location $\mathbf{a}_i$, we apply a sharp Gaussian effect with a randomly sampled amplitude $A_i \sim \mathrm{Unif}(-10, 18)$ and radius $r_i \sim \mathrm{Unif}(0.005, 0.03)$.} Letting $\mathbf{x}(\mathbf{s})$ denote the 3D Cartesian coordinates of the spherical location $\mathbf{s} = (\lambda, \theta)$, the final deterministic spatial trend is bounded below at $0.5$ and defined as:
$$m(\mathbf{s}) = \max\left\{ 0.5, f_{macro}(\lambda, \theta) + \sum_{i=1}^{60} A_i \exp \left( - \frac{\|\mathbf{x}(\mathbf{s}) - \mathbf{x}(\mathbf{a}_i)\|^2}{r_i} \right) \right\}$$

\begin{figure}[H]
    \centering
    \includegraphics[scale=0.36]{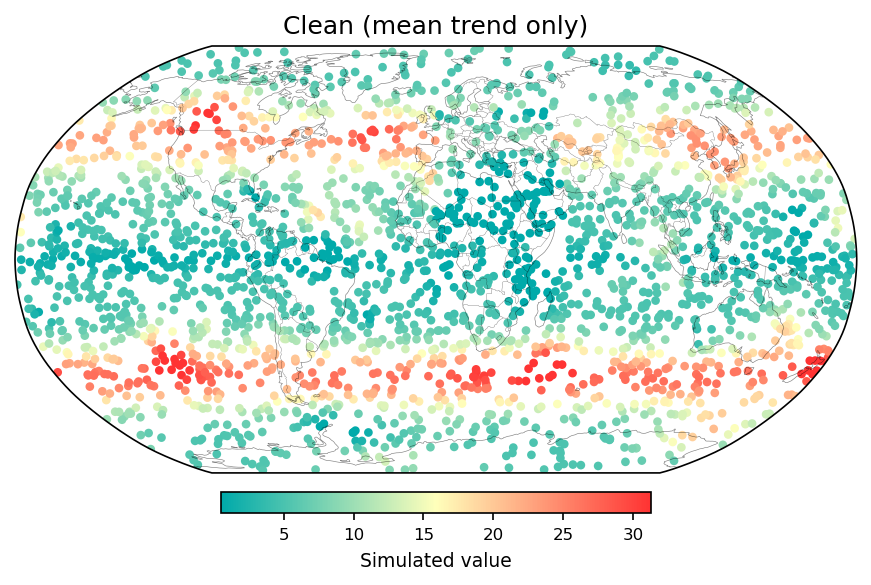}
    \includegraphics[scale=0.36]{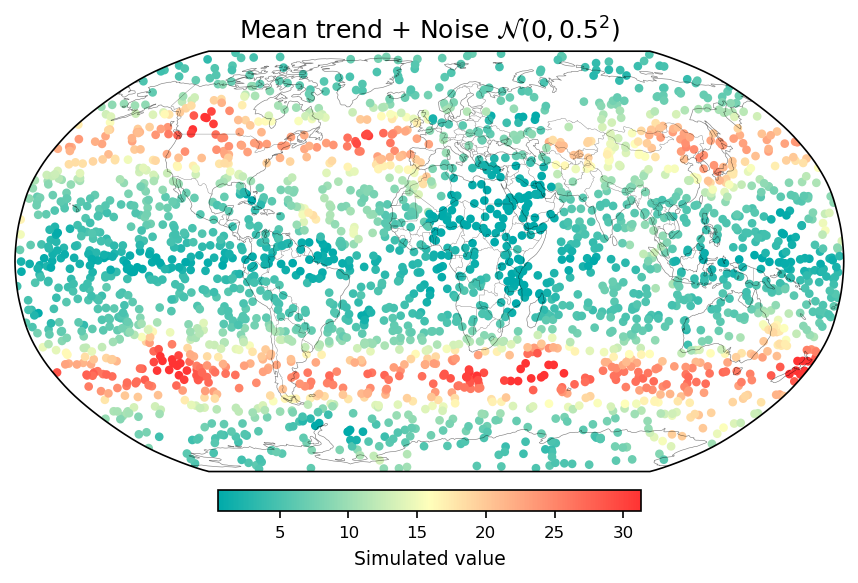}
    \includegraphics[scale=0.36]{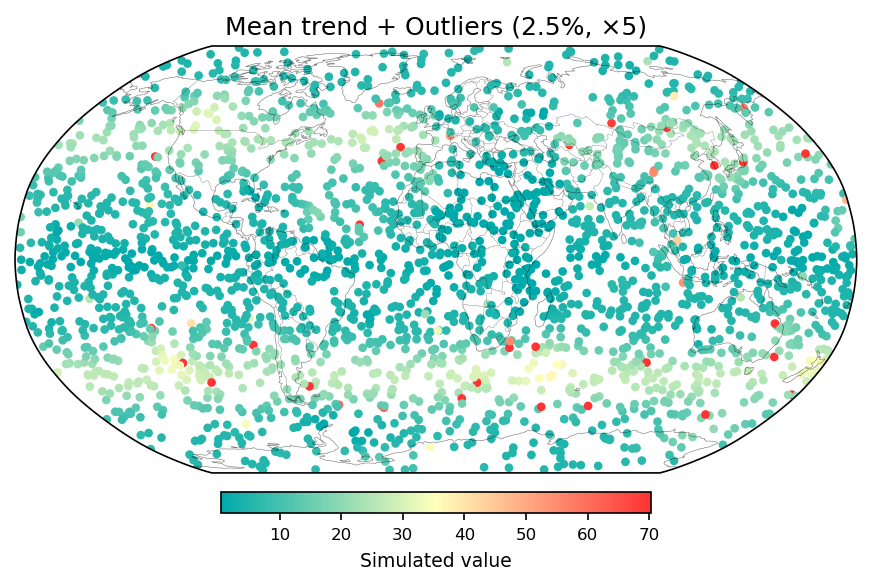}
    \caption{{Realizations of Local Extremes (ii)}}
    \label{fig:sim2}
\end{figure}

The plot of one realization is shown in Figure \ref{fig:sim2}. The prediction results are summarized in Table \ref{tab:noise_free_extremes} - Table \ref{tab:noise_free_outliers}. 
Models utilizing the Euclidean Wendland basis (OLS\_W and DK\_W) fail to adequately capture the complex deterministic structure, yielding poor predictive performance. Conversely, approaches based on the spherical MRTS basis demonstrate vastly superior approximation capabilities. Notably, when introduced to extreme artificial outliers, the robust spherical DeepKriging model trained with the Huber loss (DK\_S\_H) maintains the highest predictive accuracy (RMSE $= 7.51 \pm 2.58$), outperforming both Universal Kriging (RMSE $= 8.11 \pm 2.38$) and the standard spherical DeepKriging model (RMSE $= 8.54 \pm 2.30$). This confirms that the spherical MRTS framework, particularly when coupled with robust regularization, intrinsically excels at modeling severe, highly localized deterministic features.

\begin{table}[ht]
  \centering
  \small
  {
  \begin{tabular}{@{} lcc @{}}
    \toprule
    Model & RMSE & MAE \\
    \midrule
    OLS\_W    & $32.42 \pm 70.57$ & $7.10 \pm 4.65$ \\
    OLS\_S    & $1.38 \pm 0.41$ & $0.64 \pm 0.26$ \\
    DK\_W     & $5.63 \pm 0.44$ & $4.04 \pm 0.32$ \\
    DK\_MRTS  & $1.01 \pm 0.23$ & $0.58 \pm 0.11$ \\
    DK\_S     & $0.77 \pm 0.18$ & $0.37 \pm 0.07$ \\
    DK\_S\_H  & $0.74 \pm 0.18$ & $0.35 \pm 0.05$ \\
    UK        & $0.89 \pm 0.16$ & $0.45 \pm 0.05$ \\
    \bottomrule
  \end{tabular}
  }
  \caption{{Prediction performance for the Local Extremes dataset without $\varepsilon(\bm{s})$}}
  \label{tab:noise_free_extremes}
\end{table}

\begin{table}[H]
  \centering
  \small
  {
  \begin{tabular}{@{} lcc @{}}
    \toprule
    Model & RMSE & MAE \\
    \midrule
    OLS\_W    & $27.00 \pm 63.48$ & $6.78 \pm 4.23$ \\
    OLS\_S    & $1.64 \pm 0.27$ & $1.01 \pm 0.15$ \\
    DK\_W     & $5.67 \pm 0.42$ & $4.08 \pm 0.32$ \\
    DK\_MRTS  & $1.19 \pm 0.19$ & $0.79 \pm 0.09$ \\
    DK\_S     & $0.99 \pm 0.14$ & $0.67 \pm 0.06$ \\
    DK\_S\_H  & $1.00 \pm 0.15$ & $0.67 \pm 0.05$ \\
    UK        & $1.08 \pm 0.13$ & $0.72 \pm 0.05$ \\
    \bottomrule
  \end{tabular}
  }
  \caption{{Prediction performance for the Local Extremes dataset with $\varepsilon(\bm{s})\sim N(0,0.5^2)$}}
  \label{tab:noise_normalnoise}
\end{table}

\begin{table}[H]
  \centering
  \small
  \begin{tabular}{@{} lcc @{}}
    \toprule
    Model & RMSE & MAE \\
    \midrule
    OLS\_W    & $37.01 \pm 62.24$ & $8.67 \pm 4.27$ \\
    OLS\_S    & $8.20 \pm 2.27$ & $3.39 \pm 0.58$ \\
    DK\_W     & $10.32 \pm 2.16$ & $5.67 \pm 0.54$ \\
    DK\_MRTS  & $8.69 \pm 2.42$ & $3.53 \pm 0.70$ \\
    DK\_S     & $8.54 \pm 2.30$ & $2.85 \pm 0.87$ \\
    DK\_S\_H  & $7.51 \pm 2.58$ & $1.50 \pm 0.44$ \\
    UK        & $8.11 \pm 2.38$ & $2.85 \pm 0.46$ \\
    \bottomrule
  \end{tabular}
  \caption{{Prediction performance for the Local Extremes dataset with the artificial outliers.}}
  \label{tab:noise_free_outliers}
\end{table}

\newpage
\noindent{\textbf{(iii) Non-stationary Local Extremes Simulator on $\mathbb{S}^2$}}\vspace{0.2cm}\par

To simulate a complex and non-stationary setting, inspired from similar to Experiment 4 in \citet{lin2023some}, we begin with a Gaussian stationary process $\kappa(\mathbf{s})$ with a mean of 1 and the  exponential covariance function, where the variance and range parameters are set to $0.1$ and $1.5$, respectively. The resulting process is then transformed via the Wilson--Hilferty transformation to approximate a Gamma distribution with parameters, and subsequently centered:
\begin{align}
 {\eta(\mathbf{s})} &= \frac{a}{b} \left( 1 - \frac{1}{9a} + \kappa(\mathbf{s}) \sqrt{\frac{1}{9a}} \right)^3, \nonumber \\
 g(\mathbf{s}) &= {\eta(\mathbf{s})} - \mathbb{E}[\eta(\mathbf{s})].
\end{align}
 In the following simulation setting, we consider $a=2$ and $b=1.$
 To induce heteroscedasticity and non-stationarity, we employ the same complex deterministic mean structure $m(\mathbf{s})$ defined in (ii). The final simulated field is constructed as the sum of this mean trend and the transformed spatial component, and is truncated at zero to ensure non-negativity:
\begin{align}
 y(\mathbf{s}) = \max\bigl(m(\mathbf{s}) + g(\mathbf{s}),\, 0\bigr).
\end{align}

The plot of one realization is shown in Figure \ref{fig:sim3}.
The results for the noise-free, white-noise, and artificial outlier scenarios are presented in Tables~\ref{tab:local_extremes_c15}--\ref{tab:local_extremes_outliers_c15}, respectively.

\begin{figure}[H]
    \centering
    \includegraphics[scale=0.36]{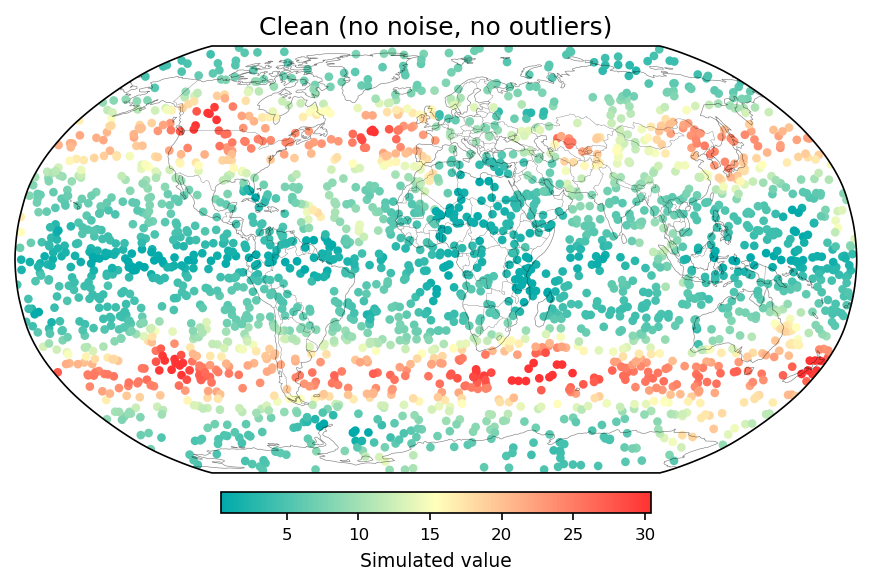}
    \includegraphics[scale=0.36]{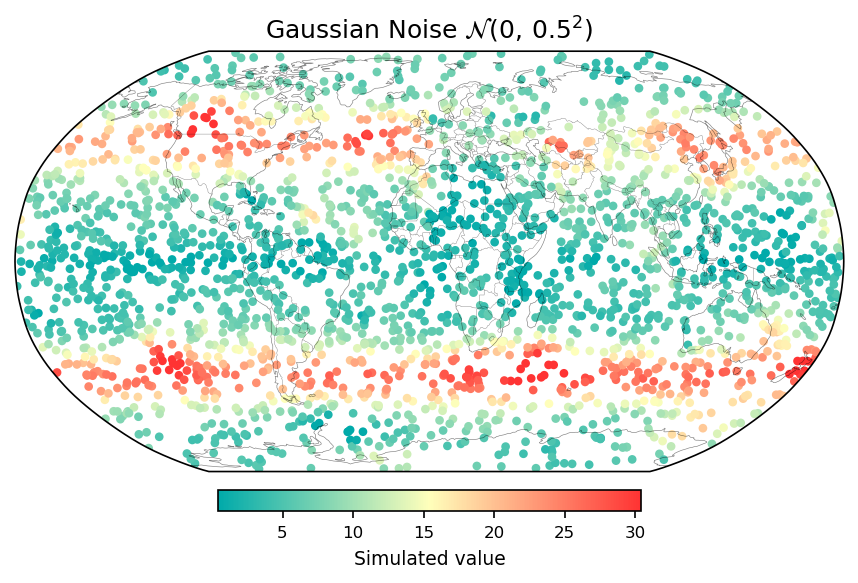}
    \includegraphics[scale=0.36]{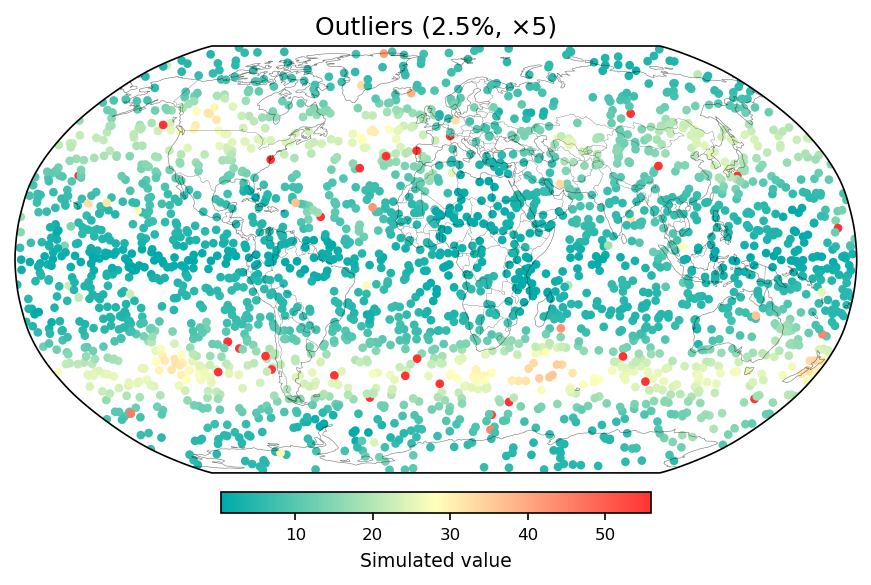}
    \caption{{Realizations of Non-stationary Local Extremes (iii)}}
    \label{fig:sim3}
\end{figure}

Consistent with the purely deterministic case (ii),  DK\_S\_H maintains the lowest RMSE and MAE in all scenarios. Compared with the conventional Euclidean Wendland basis, the use of spherical basis functions provides improved predictive performance for global data. Moreover, the predictive advantage of the proposed DeepKriging method with spherical bases is particularly pronounced when the underlying spatial process is nonstationary or when the measurement noise deviates substantially from the normality assumption.

\begin{table}[H]
  \centering
  \small
 {
  \begin{tabular}{@{} lcc @{}}
    \toprule
    Model & RMSE & MAE \\
    \midrule
    OLS\_W    & $34.44 \pm 71.42$ & $7.68 \pm 4.86$ \\
    OLS\_S    & $1.72 \pm 0.34$   & $0.99 \pm 0.28$ \\
    DK\_W     & $6.19 \pm 0.76$   & $4.41 \pm 0.52$ \\
    DK\_MRTS  & $1.25 \pm 0.26$   & $0.79 \pm 0.16$ \\
    DK\_S     & $1.01 \pm 0.24$   & $0.57 \pm 0.14$ \\
    DK\_S\_H  & $0.98 \pm 0.22$   & $0.57 \pm 0.13$ \\
    UK        & $1.10 \pm 0.21$   & $0.64 \pm 0.13$ \\
    \bottomrule
  \end{tabular}
  }
  \caption{{Prediction performance for the non-stationary Local Extremes dataset without $\varepsilon(\bm{s})$.}}
  \label{tab:local_extremes_c15}
\end{table}

\begin{table}[H]
  \centering
  \small
{
  \begin{tabular}{@{} lcc @{}}
    \toprule
    Model & RMSE & MAE \\
    \midrule
    OLS\_W    & $29.16 \pm 64.31$ & $7.34 \pm 4.50$ \\
    OLS\_S    & $1.87 \pm 0.28$   & $1.25 \pm 0.20$ \\
    DK\_W     & $6.24 \pm 0.81$   & $4.47 \pm 0.57$ \\
    DK\_MRTS  & $1.39 \pm 0.22$   & $0.95 \pm 0.13$ \\
    DK\_S     & $1.18 \pm 0.20$   & $0.82 \pm 0.12$ \\
    DK\_S\_H  & $1.18 \pm 0.19$   & $0.81 \pm 0.11$ \\
    UK        & $1.26 \pm 0.20$   & $0.85 \pm 0.12$ \\
    \bottomrule
  \end{tabular}
  }
  \caption{{Prediction performance for the non-stationary Local Extremes dataset with $\varepsilon(\bm{s})\sim N(0,0.5^2)$}}
  \label{tab:nonstat_normalnoise}
\end{table}

\begin{table}[H]
  \centering
  \small
 {
  \begin{tabular}{@{} lcc @{}}
    \toprule
    Model & RMSE & MAE \\
    \midrule
    OLS\_W    & $49.14 \pm 76.67$ & $10.12 \pm 5.68$ \\
    OLS\_S    & $9.31 \pm 2.96$ & $3.55 \pm 0.62$ \\
    DK\_W     & $11.67 \pm 3.03$ & $6.07 \pm 0.76$ \\
    DK\_MRTS  & $10.09 \pm 3.06$ & $3.83 \pm 0.77$ \\
    DK\_S     & $9.79 \pm 3.02$ & $3.12 \pm 0.77$ \\
    DK\_S\_H  & $8.85 \pm 3.19$ & $1.88 \pm 0.59$ \\
    UK        & $9.29 \pm 3.04$ & $3.10 \pm 0.65$ \\
    \bottomrule
  \end{tabular}
  }
  \caption{{Prediction performance for the non-stationary Local Extremes dataset with the artificial outliers.}}
  \label{tab:local_extremes_outliers_c15}
\end{table}


\subsection{Real Data}
To demonstrate the practical predictive performance of our method, we conduct experiments on real-world data from the MERRA-2 Surface Flux Diagnostics dataset, managed by the NASA Goddard Earth Sciences Data and Information Services Center (\url{https://disc.gsfc.nasa.gov/datasets/M2T1NXFLX_5.12.4/summary?keywords=flx_Nx}). From this dataset, we extract three distinct types of global spatial variables—surface temperature, and wind speed—measured on January 1, 2024, at a spatial resolution of $0.5^\circ$ latitude $\times$ $0.625^\circ$ longitude. These variables exhibit different spatial dependence structures and levels of variability, providing a comprehensive benchmark for evaluating the proposed approach.
In the following studies, we randomly select one hourly observation as the experimental subset. From this selected hour, we further randomly sample 200,000 spatial locations for analysis. The data are then divided into 80\% for training, 10\% for validation, and 10\% for testing. To ensure the robustness of our results, all experiments are repeated ten times.

\paragraph{(a) Temperature Dataset}\mbox{}\par
The temperature dataset (Figure \ref{fig:tem}) typically displays the smoothest spatial behavior among the three variables. Global temperature fields are largely driven by broad-scale climatic patterns, such as latitudinal gradients, seasonal cycles, and large-scale ocean–atmosphere interactions. As a result, temperature varies gradually over space and exhibits strong spatial continuity across wide regions. Extreme temperature values may occur, but they are usually associated with coherent large-scale phenomena rather than highly localized spikes. Consequently, temperature serves as a representative example of a relatively smooth global process that can often be well captured by models emphasizing large-scale dependence.

\begin{table}[H]
  \centering
  \small
{
  \begin{tabular}{@{} lcccc @{}}
    \toprule
    \textbf{Model}  & \textbf{RMSE} & \textbf{MAE} \\
    \midrule
    OLS\_W      & $16.29 \pm 0.06$ & $12.56 \pm 0.06$  \\
    OLS\_S     & $1.78 \pm 0.02$  & $1.08 \pm 0.01$   \\
    DK\_W        & $13.58 \pm 0.10$ & $8.77 \pm 0.13$  \\
    DK\_MRTS    & $0.56 \pm 0.02$  & $0.33 \pm 0.01$   \\
    DK\_S      & $0.45 \pm 0.08$  & $0.24 \pm 0.06$   \\
    DK\_S\_H    & $0.49 \pm 0.09$  & $0.27 \pm 0.06$  \\
    UK         & $0.48 \pm 0.02$  & $0.23 \pm 0.00$   \\
    \bottomrule
  \end{tabular}
  }
  \caption{{Prediction comparison under the temperature dataset.}}
\end{table}

\paragraph{(b) Windspeed Dataset}\mbox{}\par

Compared with Temperature Dataset, the wind speed dataset (Figure \ref{fig:wind}) exhibits highly heterogeneous and non-smooth spatial structures. Its large-scale variability is primarily driven by atmospheric circulation systems, including trade winds, jet streams, and storm tracks. Consequently, wind speed fields often display coherent regional patterns with gradual large-scale spatial transitions, while simultaneously exhibiting patchy spatial distributions and sharp local gradients, making wind speed substantially more complex and challenging to predict than temperature fields.

\begin{table}[H]
  \centering
  \small
{
  \begin{tabular}{@{} lcccc @{}}
    \toprule
    \textbf{Model}  & \textbf{RMSE} & \textbf{MAE} \\
    \midrule
    OLS\_W     & $3.56 \pm 0.02$ & $2.66 \pm 0.01$  \\
    OLS\_S     & $1.59 \pm 0.02$ & $1.13 \pm 0.01$  \\
    DK\_W        & $3.22 \pm 0.05$ & $2.26 \pm 0.05$  \\
    DK\_MRTS    & $0.45 \pm 0.01$ & $0.29 \pm 0.01$  \\
    DK\_S      & $0.31 \pm 0.01$ & $0.17 \pm 0.01$  \\
    DK\_S\_H   & $0.32 \pm 0.01$ & $0.17 \pm 0.00$  \\
    UK         & $0.39 \pm 0.01$ & $0.21 \pm 0.00$  \\
    \bottomrule
  \end{tabular}
  }
  \caption{{Prediction comparison under windspeed dataset.}}
\end{table}

In summary, across diverse types of spatial processes, DeepKriging models equipped with the spherical MRTS basis consistently demonstrate superior predictive performance. This advantage is particularly pronounced for highly heterogeneous and non-smooth data, such as wind speed, where localized extremes and strong spatial irregularities pose substantial challenges for conventional basis constructions.

\section{Conclusion}\label{sec:conclusion}
This paper develops a DeepKriging framework for spatial prediction on spherical domains by integrating spherical multi-resolution thin-plate spline (MRTS) basis functions into the neural network input layer. Motivated by the limitations of Euclidean basis constructions for global data, the proposed approach defines basis features intrinsically on $\mathbb{S}^2$, thereby respecting spherical geometry and providing a flexible multi-resolution representation ordered by Laplace--Beltrami roughness.

Comprehensive simulation studies demonstrate that the proposed spherical-basis DeepKriging approach delivers improved predictive accuracy relative to benchmark methods, including OLS, universal kriging, and DeepKriging models built on Euclidean Wendland and Euclidean MRTS bases. The gains are especially pronounced when the latent process exhibits nonstationary structure and when the observation noise departs from Gaussianity, including heavy-tailed contamination and artificial outliers. Incorporating the Huber loss further enhances robustness under contaminated settings.

We also evaluate the proposed method on large-scale global datasets from the MERRA-2 Surface Flux Diagnostics product, using temperature and wind speed fields as representative variables with increasing degrees of spatial complexity. Across both datasets,  DeepKriging with spherical MRTS features consistently achieves strong predictive performance, with particularly clear advantages for wind speed, where spatial heterogeneity, intermittency, and localized extremes pose substantial challenges for conventional basis constructions.

These results highlight the importance of geometry-aware basis representations in deep spatial prediction for global data. Future work includes extending the framework to spatio-temporal settings and developing uncertainty quantification procedures tailored to spherical DeepKriging for extreme-event prediction.

\begin{figure}[H]
    \centering
    \includegraphics[scale=0.28]{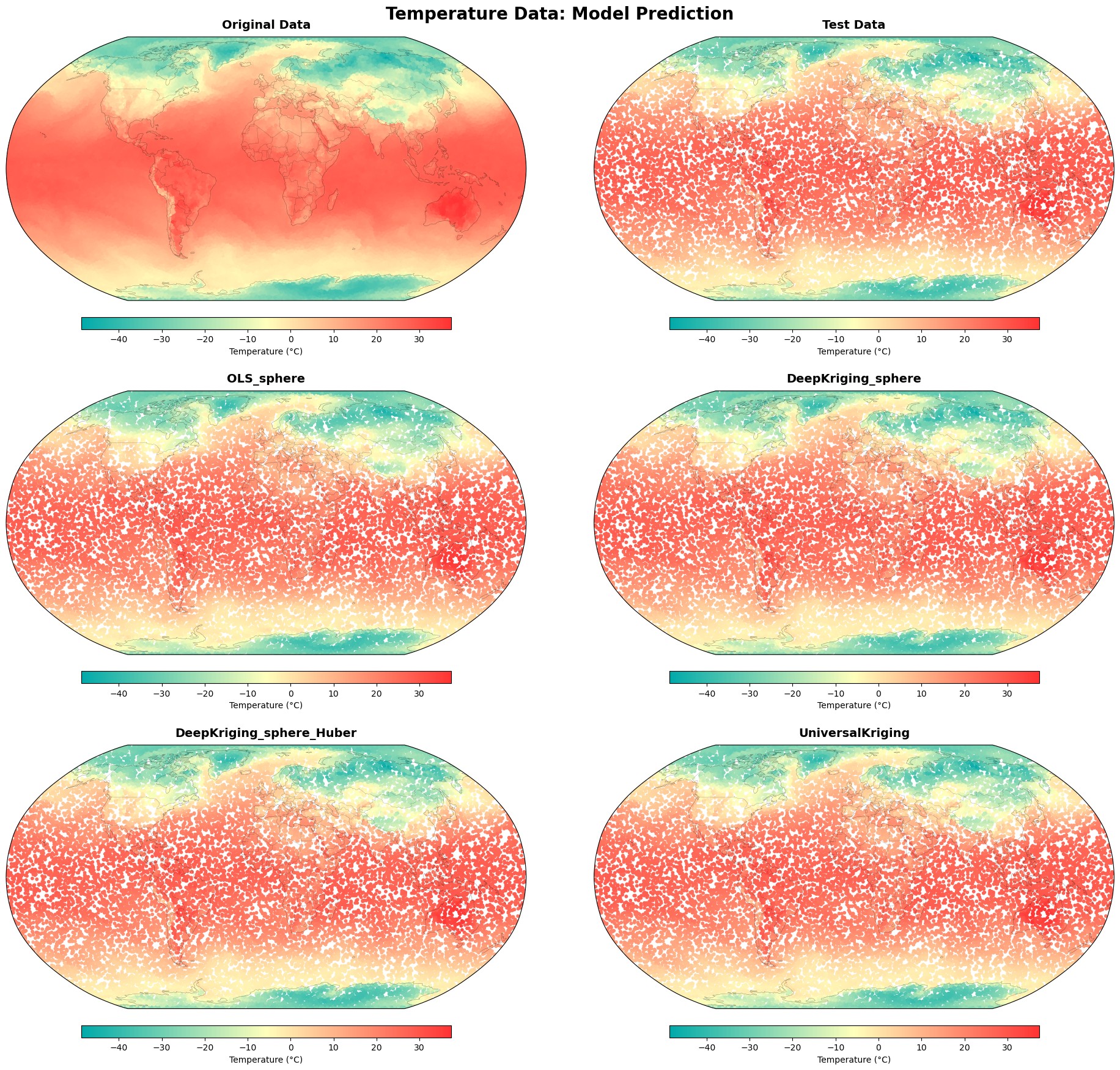}
    \includegraphics[scale=0.28]{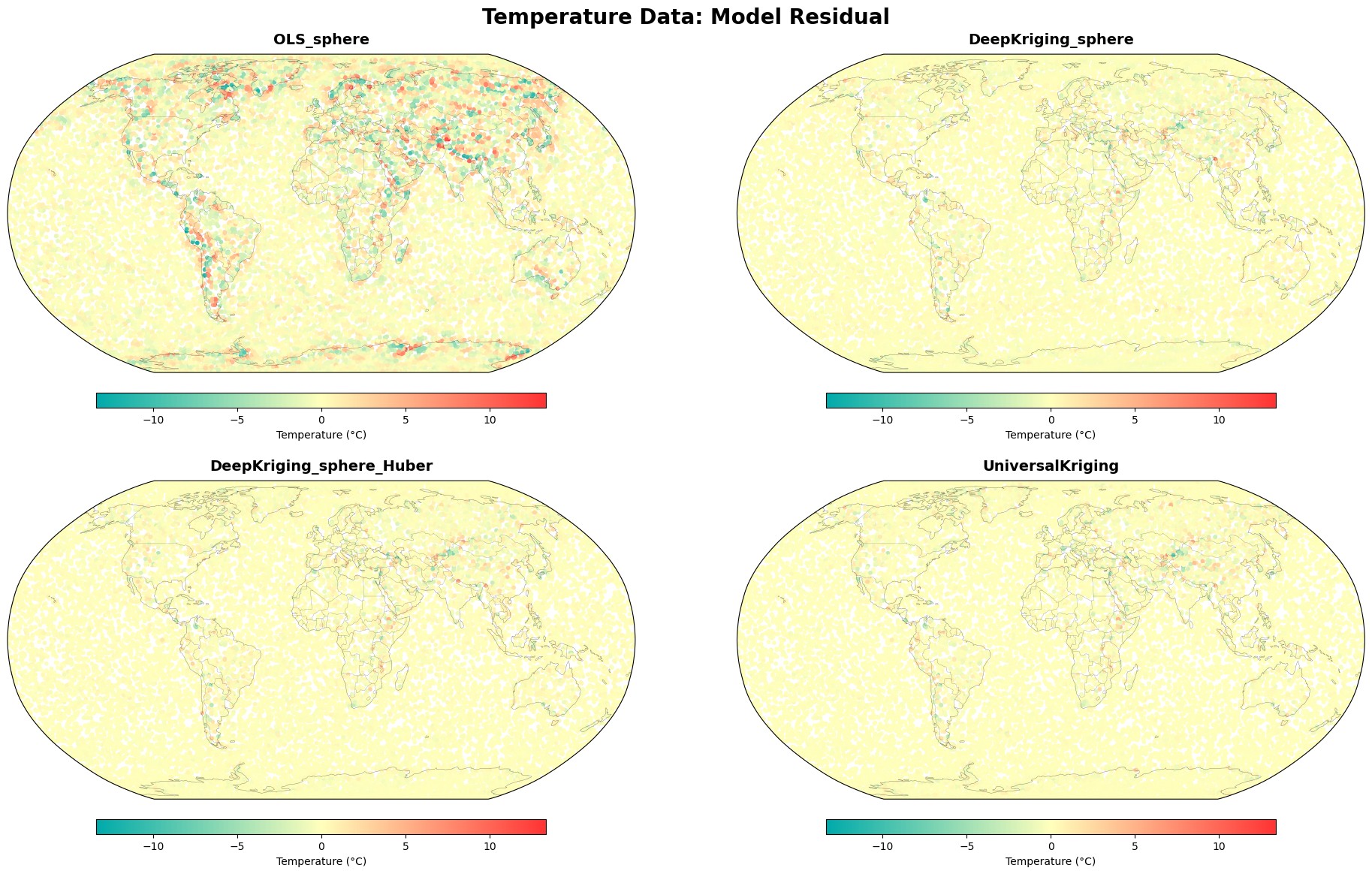}
    \caption{Temperature data model prediction and residual.}
    \label{fig:tem}
\end{figure}

\begin{figure}[H]
    \centering
    \includegraphics[scale=0.28]{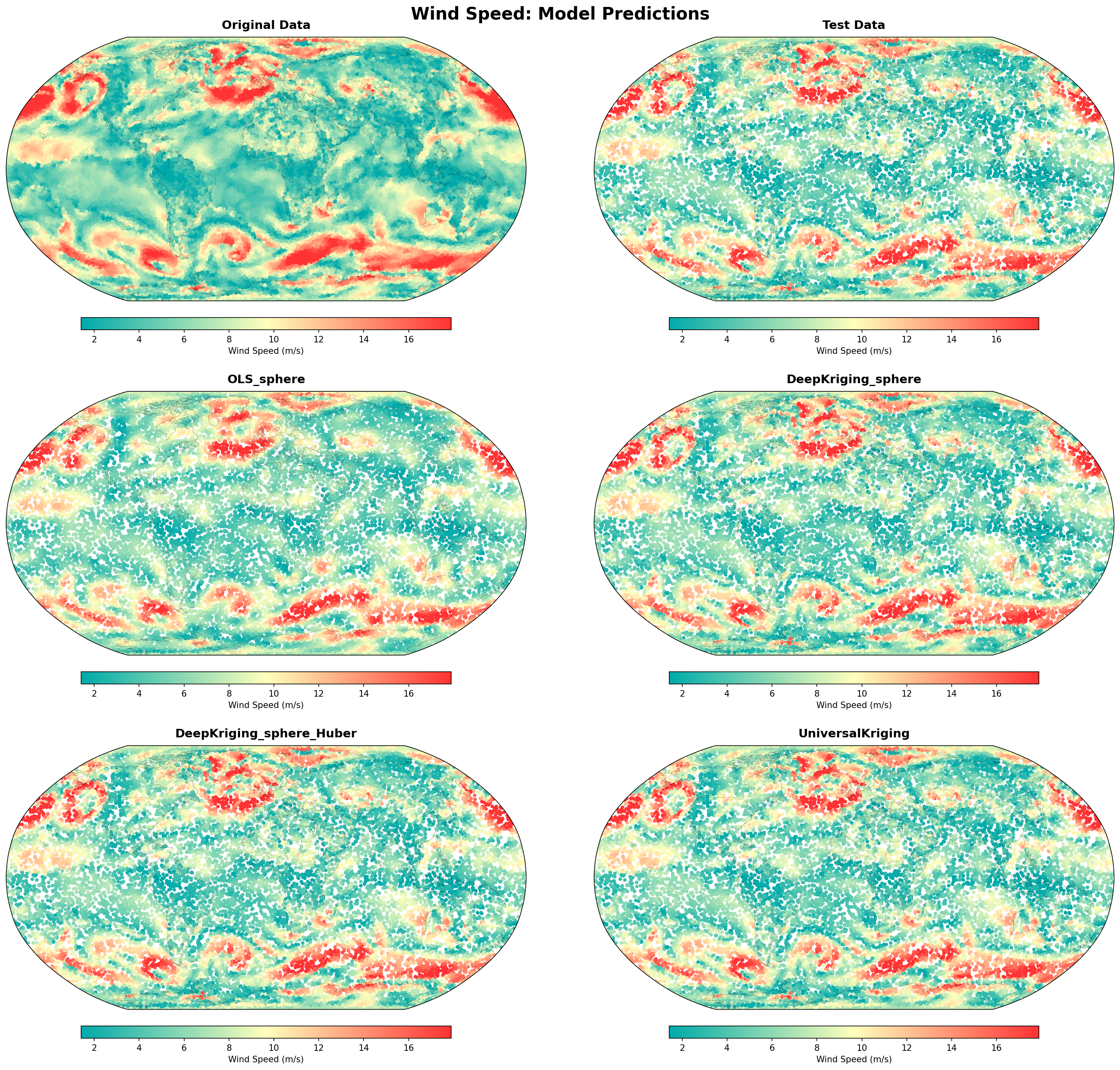}
    \includegraphics[scale=0.28]{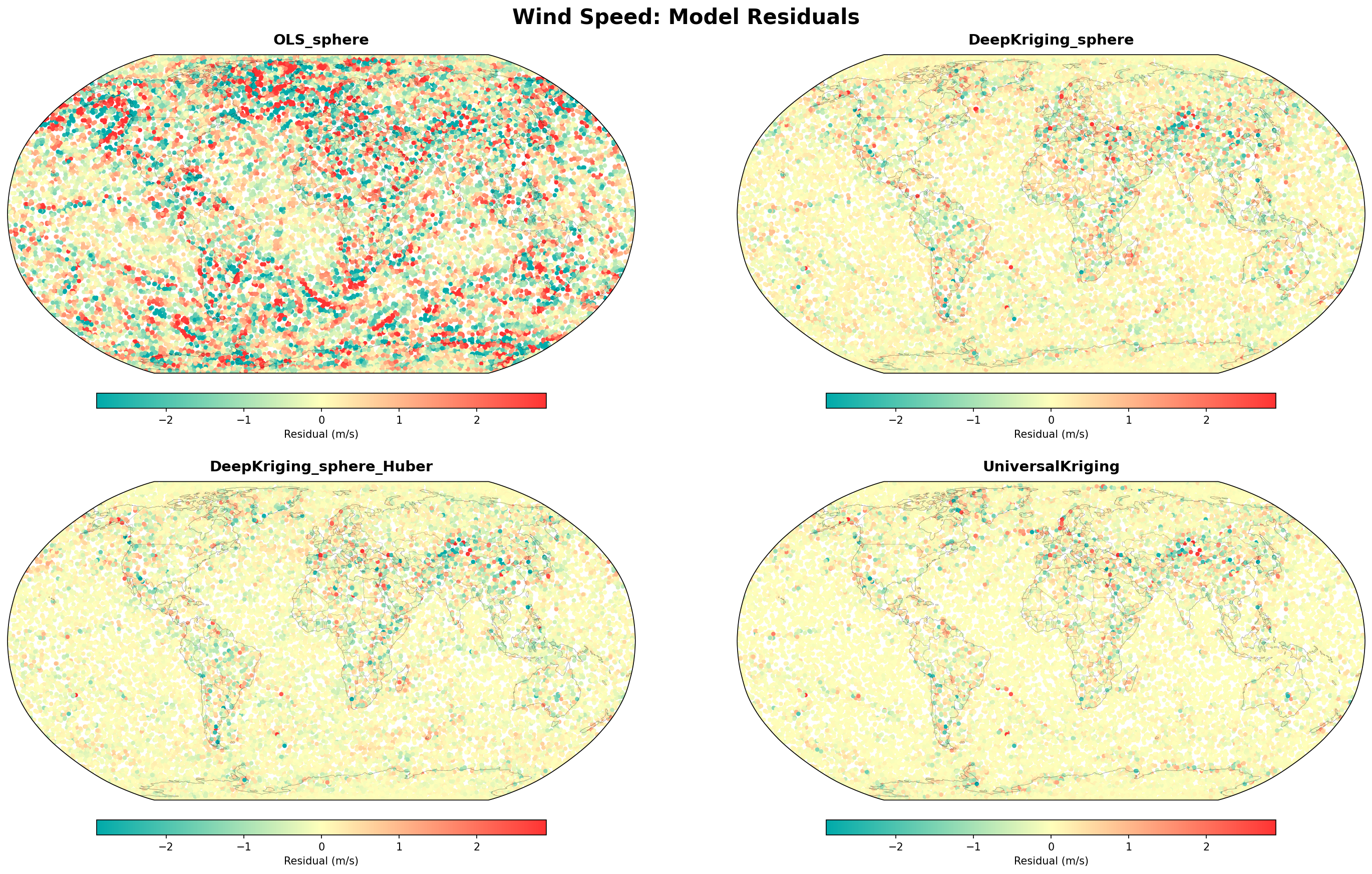}
    \caption{Windspeed data model prediction and residual.}
    \label{fig:wind}
\end{figure}


\bibliographystyle{chicago}
\bibliography{main_refwu}
\end{document}